\pdfoutput=1
\input{aipcheck}

\documentclass[
    final            % use final for the camera ready runs
  ]
  {aipproc}

\layoutstyle{6x9}

\newcommand{\nc}{\newcommand}
\nc{\bc}{\begin{center}}
\nc{\ec}{\end{center}}
\nc{\mb}[1]{\makebox[#1]{}}
\nc{\CC}{{\scriptscriptstyle CC}}
\nc{\NC}{{\scriptscriptstyle NC}}
\nc{\CHPT}{{$\chi_{\PT}$\ }}
\nc{\st}{\scriptstyle}
\nc{\sst}{\scriptscriptstyle}
\nc{\mco}{\multicolumn}
\nc{\epp}{\epsilon^{\prime}}
\nc{\vep}{\varepsilon}
\nc{\ra}{\rightarrow}
\nc{\ppg}{\pi^+\pi^-\gamma}
\nc{\nuN}{{\nu N_0}}
\nc{\nub}{{\overline{\nu}}}
\nc{\nubN}{{\nub N_0}}
\nc{\snuNC}{{\langle \sigma^{\nuN}_{\NC}\rangle }}
\nc{\snubNC}{{\langle \sigma^{\nubN}_{\NC}\rangle }}
\nc{\snuCC}{{\langle \sigma^{\nuN}_{\CC}\rangle }}
\nc{\snubCC}{{\langle \sigma^{\nubN}_{\CC}\rangle }}
\nc{\Rnu}{{R^{\nu}}}
\nc{\Rnub}{{R^{\overline{\nu}}}}
\nc{\sintW}{{\sin^2 \theta_{W} }}
\nc{\vp}{{\bf p}}
\nc{\rz}{{\rho_0^2}}
\nc{\ko}{K^0}
\nc{\kb}{\bar{K^0}}
\nc{\al}{\alpha}
\nc{\ab}{\bar{\alpha}}
\newcommand{\be}{\begin{equation}}
\newcommand{\ee}{\end{equation}}
\newcommand{\bea}{\begin{eqnarray}}
\newcommand{\eea}{\end{eqnarray}}

\usepackage{graphicx}
\usepackage{subfigure}

\begin{document}

\title{The Determination of $\sintW$ in Neutrino Scattering: no more anomaly}


\author{A.~W.~Thomas}{
address={ARC Centre of Excellence in Particle Physics at the Tera-scale 
(CoEPP) and \\
CSSM, School of Chemistry and Physics,  
University of Adelaide, Adelaide SA 5005, Australia}
}

\begin{abstract}
We review the corrections to the NuTeV determination of $\sintW$, 
concluding that it is no longer appropriate to present it as an 
``anomaly''. Indeed, when well understood corrections associated 
with charge symmetry violation and the iso-vector nuclear force 
are properly included, the measurement is 
completely consistent with the 
Standard Model.
\end{abstract}

\maketitle

\section{Introduction}
The famous expression of Paschos and Wolfenstein~\cite{Paschos:1972kj} 
relating the ratio of neutral-current and charge-changing neutrino
interactions on isoscalar targets to the Weinberg angle takes the form:
\be
 R^- \equiv { \rz \left( \snuNC - \snubNC \right) \over
 \snuCC - \snubCC } = \frac{1}{2} - \sintW  .
\label{eq:PasW}
\ee
In Eq.(\ref{eq:PasW}), $\snuNC$ and $\snuCC$ are respectively the
neutral-current and charged-current inclusive, total
cross sections for neutrinos on an
isoscalar target (assuming charge symmetry is exact and that there
is no strange quark asymmetry).  
The quantity $\rho_0 \equiv M_{W}/(M_{Z}\,\cos
\theta_{W})$ is one in the Standard Model.

The application of this relation in a test of the Standard Model took 
almost 30 years, with the 
NuTeV collaboration reporting the measurement of 
neutrino charged-current and
neutral-current cross sections on iron \cite{Zeller:2001hh}. 
Surprisingly, the result that they found, namely 
$\sintW = 0.2277 \pm 0.0013~({\rm stat}) \pm 0.0009 ~({\rm syst})$, 
differed from the expectation within the Standard Model by $3 \sigma$.
This has been widely treated as an indication of the need to go beyond 
the Standard Model.

Unfortunately, there are several corrections to the Paschos-Wolfenstein 
relation (PWR) that were omitted in the NuTeV analysis~\cite{Davidson:2001ji}. 
The first of these arises because  
in the Standard Model the $u$ and $d$ masses 
are not equal and so charge symmetry, a necessary condition for the 
PWR, is broken~\cite{Londergan:2009kj,Londergan:1998ai}. 
Indeed, two independent estimates of the effect of charge 
symmetry violation in parton distributions, which had been published a 
decade before NuTeV~\cite{Sather:1991je,Rodionov:cg} 
and which reduce the discrepancy to $2 \sigma$ or less, 
were ignored in the NuTeV analysis. The dubious response to that criticism 
has been that ``there were no reliable calculations''. We address that 
claim in the next section, showing that it is clearly incorrect.

Of course, the steel target used by NuTeV is not iso-scalar and 
standard corrections were made for the neutron excess. 
However, the second major correction to the NuTeV analysis involves a 
critical piece of physics which was {\it not} appreciated at the 
time. That is, the EMC effect produced by those extra neutrons on 
{\it every} nucleon in the nucleus 
and which is therefore not  taken into 
account by subtracting their contribution 
to the nuclear structure function.  
Because the iso-vector nuclear force is repulsive between neutrons 
and $d$-quarks, the effect of this new nuclear correction is to 
shift momentum from {\it all} $u$ to {\it all} $d$ quarks 
in the nucleus~\cite{Cloet:2009qs}. 
Thus, as far as the determination of $\sintW$ goes, it has the same sign 
and similar magnitude to the CSV correction already mentioned.

In this brief review we first outline the reasons why the original 
calculation of the size of CSV was considered to be very reliable by 
the authors. We then describe the recent direct measurement of this 
effect within the framework of lattice QCD, which agreed beautifully 
with the earlier bag model based caclulations. Next we explain the 
role of the iso-vector EMC effect. Finally, we make some 
concluding remarks.

\section{Model Independence of the Charge Symmetry Correction}
The first calculations of CSV in parton
distributions were made independently by Sather and 
Rodionov {\it et al.}~\cite{Sather:1991je,Rodionov:cg}. 
Both were based on a calculation of the parton distribution 
functions (PDFs)
at a low momentum scale, appropriate to a valence-dominated quark model,
and followed by QCD evolution to generate the CSV
distributions at the $Q^2$
values appropriate for the NuTeV experiment.
However, the two calculations were based on very different levels of 
approximation. The first~\cite{Sather:1991je} used several approximations 
to simplify the evaluation and which have the effect that the model 
distributions do not have the correct support. These approximations 
were not necessary in the work of 
Rodionov~{\it et al.}~\cite{Rodionov:cg}, in which energy-momentum 
conservation was ensured. Nevertheless, the two calculations agreed on 
the extent of CSV rather well, with the 
result a correction to the
NuTeV result $\Delta R_{CSV} \sim -0.0015$. This reduces the
reported effect from 3 to 2 standard deviations. After this was 
pointed out, NuTeV
made their own estimate of the CSV parton distributions, using a rather
different procedure~\cite{NuTeV2}. They obtained
a much smaller correction, $\Delta R_{CSV} \sim +0.0001$.
On the basis of the large discrepancy between these two results 
it was suggested that the
CSV correction might be strongly model dependent.

This question was investigated by Londergan and Thomas
\cite{Londergan:2003ij}, in order to check just how model dependent 
the size of the CSV correction could be. We briefly recall their 
argument.
The charge symmetry violating contribution to the Paschos-Wolftenstein ratio
has the form
\be
 \Delta R_{CSV} = \left[ 3\Delta_u^2 + \Delta_d^2 + \frac{4\alpha_s}{9\pi}
 \left(\bar{g}_L^2 - \bar{g}_R^2 \right) \right] \,\left[ \frac{\delta U_{V} -
 \delta D_{V}}{2(U_{V} + D_{V}) }\right]
\label{eq:CSV}
\ee
where
\bea
 \delta Q_{V} &=& \int_0^1 \, x\,\delta q_{V}(x) \,dx \nonumber \\
 \delta d_{V}(x) &=& d_{V}^p(x)- u_{V}^n(x)~; \hspace{1.0cm}
 \delta u_{V}(x) = u_{V}^p(x)- d_{V}^n(x) ~.
 \label{eq:CSVadd}
\eea
The denominator in the final term in Eq.~(\ref{eq:CSV}) gives the total
momentum carried by up and down valence quarks, while the numerator
gives the charge symmetry violating momentum
difference -- for example, $\delta U_{V}$, is the total momentum carried by up
quarks in the proton minus the momentum of down quarks in the neutron.
This ratio is completely independent of $Q^2$ and can be evaluated at
any convenient value.

Using the analytic approximation to the charge symmetry violating valence
parton distributions initially proposed
by Sather~\cite{Sather:1991je},
one can evaluate Eq.(\ref{eq:CSV}) at a low scale,
$Q_0^2$, appropriate for a (valence dominated) quark or bag
model~\cite{Signal:yc,Schreiber:tc}.
The advance over earlier work was to realize
that for NuTeV one needs only the first moments of the CSV distribution
functions and these can be obtained analytically. The result for
the moment of the CSV down valence distribution, $\delta D_{V}$, is
\bea
 \delta D_{V} &=& \int_0^1 \, x  \left[ -\frac{\delta M}{M} \frac{d}{dx}
 (x d_{V}(x)) - \frac{\delta m}{M} \frac{d}{dx} d_{V}(x) \right]
 \, dx \nonumber \\ &=& \frac{\delta M}{M} \int_0^1 \, x \, d_{V}(x)\, dx +
 \frac{\delta m}{M} \int_0^1 \, d_{V}(x)\, dx = \frac{\delta M}{M} D_{V}
 + \frac{\delta m}{M} \, ,
\label{eq:intDv}
\eea
while for the up quark CSV distribution it is
\bea
\delta U_{V} &=& \frac{\delta M}{M} \left[ \int_0^1 \, x\, \left( -
\frac{d}{dx}\left[ x u_{V}(x)\right] + \frac{d}{dx} u_{V}(x) \right)
\, dx \right] \nonumber \\
&=& \frac{\delta M}{M} \left( \int_0^1 \,x \, u_{V}(x)\, dx - \int_0^1 \,
u_{V}(x)\, dx \right) = \frac{\delta M}{M} \left( U_{V} - 2 \right)
\, .
\label{eq:intUv}
\eea
(Here $\delta M = 1.3$ MeV is the neutron-proton mass difference, and
$\delta m = m_d - m_u \sim 4$ MeV is the down-up quark mass difference.)

Equations~(\ref{eq:intDv}) and (\ref{eq:intUv}) show that
the CSV correction to the Paschos-Wolfenstein ratio depends only on
the fraction of the nucleon momentum carried by up and down valence quarks.
At no point do we have to calculate specific CSV distributions.
At the bag model scale, $Q_0^2 \approx 0.5$ GeV$^2$, the momentum fraction
carried by down valence quarks,
$D_{V}$, is between $0.2-0.33$, and the total momentum fraction carried
by valence quarks is $U_{V} + D_{V} \sim .80$.  From Eqs. (\ref{eq:intDv})
and (\ref{eq:intUv}) this gives $\delta D_{V} \approx 0.0046$,
$\delta U_{V} \approx -0.0020$.
Consequently, evaluated at the quark model scale, the CSV correction to the
Paschos-Wolfenstein ratio is
\be
 \Delta R_{CSV} \approx 0.5 \left[ 3\Delta_u^2 + \Delta_d^2 \right]
 \,\frac{\delta U_{V} - \delta D_{V}}{2(U_{V} + D_{V}) } \approx
 -0.0020 \, .
\label{eq:CSVnut}
\ee
Once the CSV correction has been calculated at some quark model scale,
$Q^2_0$, the ratio appearing in Eq.~(\ref{eq:CSV}) is independent of
$Q^2$, because both the numerator and
denominator involve the same moment of a non-singlet distribution.

Note that both Eqs.~(\ref{eq:intDv}) and (\ref{eq:intUv}) are only
weakly dependent on the choice of quark model scale -- through the
momentum fractions $D_{V}$ and $U_{V}$, which are slowly varying functions
of $Q^2_0$ and, in any case, not the dominant terms in
those equations. This, together with the $Q^2$-independence of the
Paschos-Wolfenstein ratio (Eq.~(\ref{eq:CSV})) under QCD
evolution, explains why the previous results, obtained by Londergan and
Thomas with different models and at different
values of $Q^2$~\cite{Londergan:2003pq},
were so similar. Finally, Londergan and Thomas also demonstrated that
the acceptance function calulated by NuTeV does not introduce any
significant model dependence to this result.

In spite of the power of this demonstration the effect of CSV is still 
omitted in the official NuTeV collaboration analysis of $\sintW$.

\subsection{Lattice QCD}
\begin{figure}
%\begin{center}
%\vspace*{-3mm}
\includegraphics[width=5.6in]{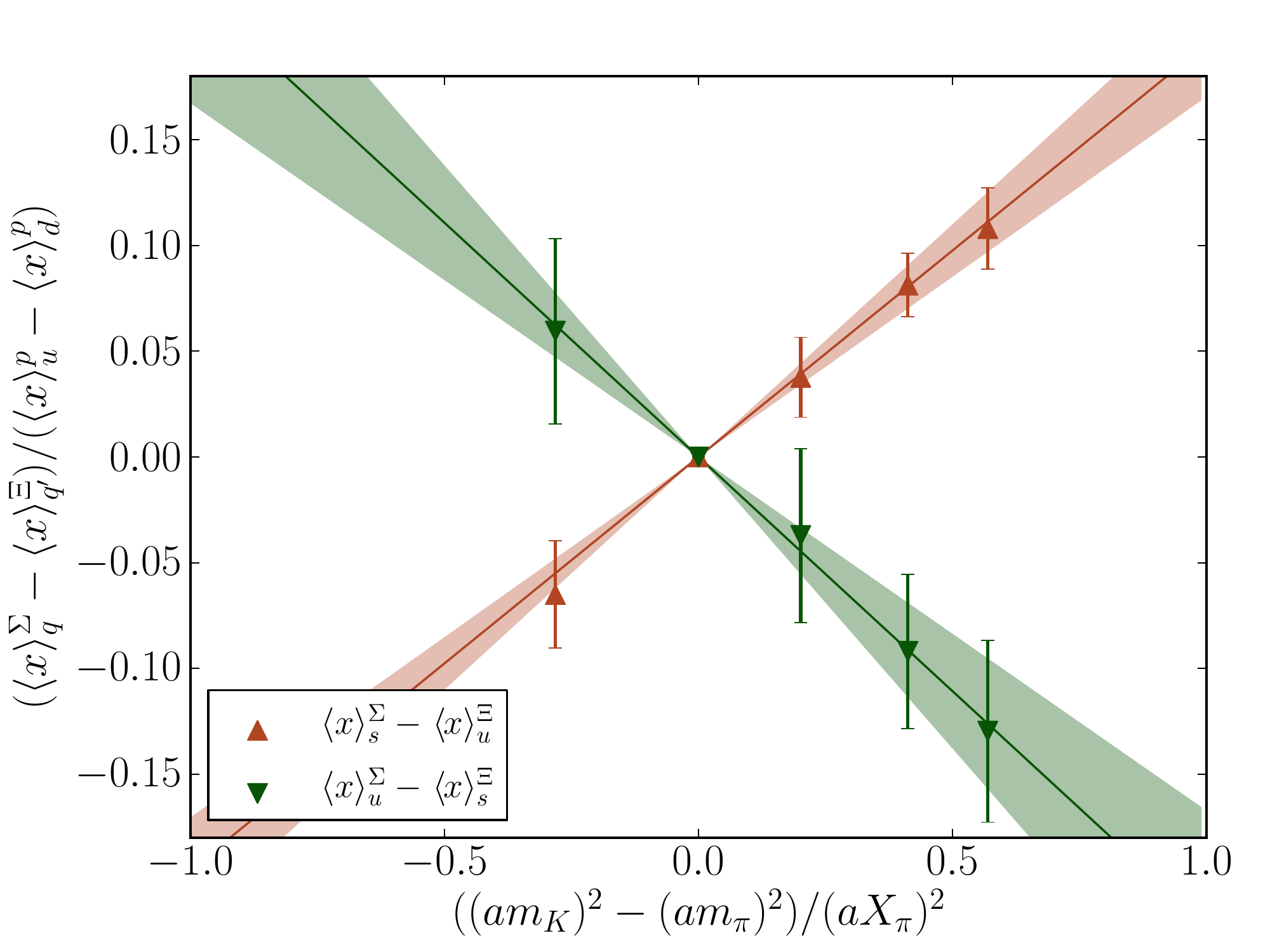}
\caption{The difference between the double and singly represented
  quarks in the $\Sigma$ and $\Xi$ as a function of the strange/light
  quark mass difference. The CSV moments $\delta u$ and $\delta d$ were 
deduced from the slopes of these lines in Ref.~\cite{Horsley:2010th}.}
\label{fig:xud}
%\vspace*{-3mm}
%\end{center}
\end{figure}

Recent simulations of PDFs within lattice QCD have included not just 
the nucleon but all members of the baryon octet, over a range of values 
of the strange quark mass. Using SU(3) symmetry one can use these results 
to investigate CSV in a novel way. Indeed, by viewing the s-quark as 
effectively a heavier light quark, one can extract the moments of the CSV 
PDFs. This is illustrated in Fig.~\ref{fig:xud}, where we show the linear 
dependence of the second moment of the CSV $u$ and $d$ distributions 
as a function of the $d-u$ mass difference. 
{}From the slopes of these lines Horsley {\it et al.} deduced the first 
moments of the $C$-positive CSV moments $\delta u = -0.0023 \pm 0.0006$ 
and $\delta d = 0.0020 \pm 0.0003$~\cite{Horsley:2010th}, 
in excellent agreement with 
the values $\delta u^- = -0.0014$ and $\delta d^- = 0.0015$ (at 
4 GeV$^2$) found within the MIT bag 
model~\cite{Sather:1991je,Rodionov:cg}.

In summary, the explicit lattice simulations of the extent of charge symmetry 
violation arising through the $u-d$ mass difference is in remarkable 
agreement with the earlier calculations based on the MIT bag model. 
Given the demonstration of the degree of model independence of the results 
for the moment relevant to the NuTeV experiment this should not be a surprise. 
It certainly provides a convincing counter to the suggestions that there have 
been no reliable calcuations. This level of agreement between lattice QCD 
simulation and phenomenology can leave no doubt concerning the sign 
and magnitude of the CSV correction which must therefore be included in any 
serious analysis of the NuTeV data.

\section{Hadrons in-medium}
A remarkably efffective approach to the nuclear many body problem, based 
on the underlying quark structure of hadrons, begins with the realization 
that at some density 
(perhaps 3 to 5 times nuclear matter density) nuclear matter will
make a transition to quark matter -- a phase transition which may have dramatic 
effects on the observable properties of neutron stars.  
One therefore constructs a theory of 
the nuclear many-body system starting with a 
description of hadron structure at the quark level and considers the 
self-consistent modification of that structure in a nuclear medium.
This is the approach taken within the QMC (quark-meson coupling) 
model~\cite{Guichon:1995ue,Saito:2005rv}. A remarkable
advantage of this approach is that no new parameters are 
needed to calculate the effective
density dependent forces~\cite{Guichon:2006er} between any hadrons 
whose quark structure is known. Indeed, it has
been possible to develop a remarkably successful derivation of realistic 
Skyrme forces~\cite{Guichon:2006er,Guichon:2004xg}
for comparison with low energy nuclear phenomenology -- 
while the fully relativistic
underlying theory successfully predicts key features of 
hypernuclear physics and allows the
study of the appearance of hyperons in dense matter~\cite{Guichon:2008zz}. 
\begin{figure}
\includegraphics[width=5.6in]{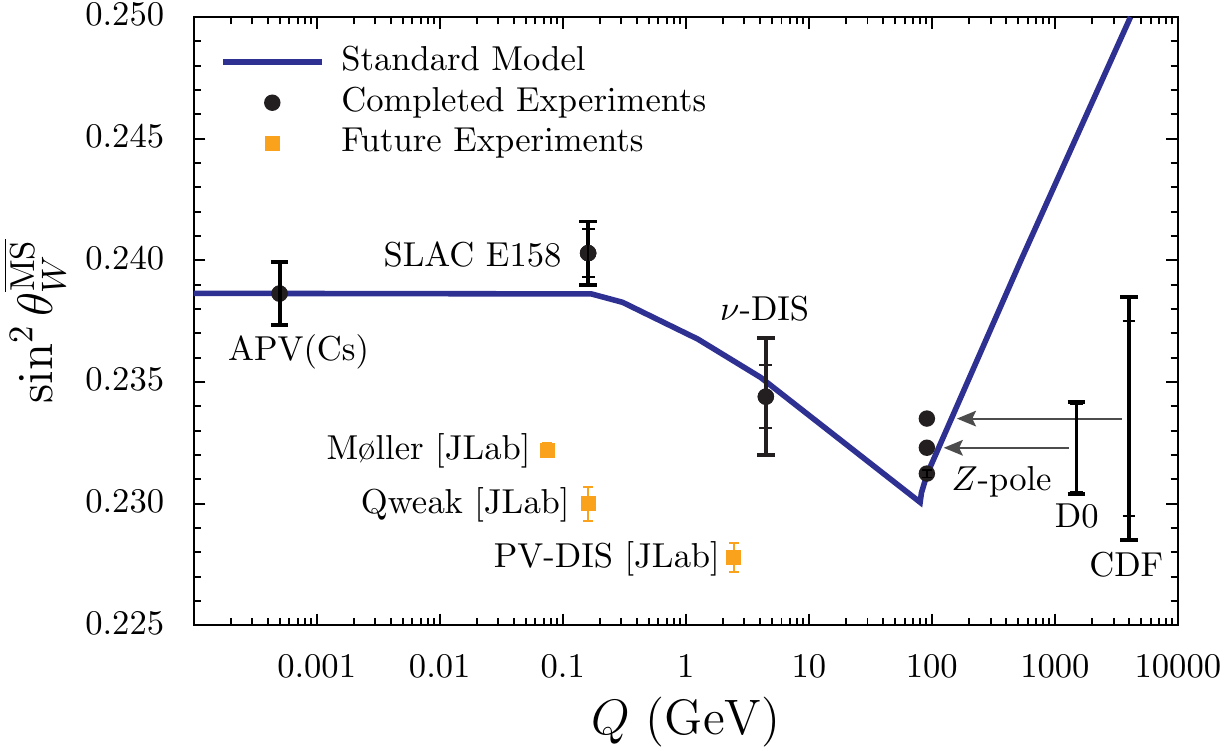}
\caption{
The curve represents the running of $\sin^2\theta_W$ in the $\bar{MS}$
renormalization scheme. 
The $Z$-pole point represents the combined results of six LEP and
SLC experiments, while the  
CDF and D0 collaboration results
(at the $Z$-pole) and the SLAC E158 result,
are labelled accordingly.
The atomic parity violating (APV) result
has been shifted from $Q^2 = 0$ for clarity.
The inner error bars represent the statistical uncertainty
and the outer error bars the total uncertainty
-- see Ref.~\cite{Bentz:2009yy} for details and associated references.
\label{fig:NuTeV}}
\end{figure}

The QMC model has the additional advantage that one 
can address not only those low energy
properties such as binding energies and 
charge densities but it can also be used to 
calculate the nuclear modification of 
the form factors~\cite{Lu:1998tn} and 
deep inelastic structure functions~\cite{Saito:1992rm}. 
Extensions of the QMC approach based upon a covariant, 
confining version of the NJL model  
have produced a satisfactory description of the EMC effect in finite 
nuclei~\cite{Cloet:2006bq}. In the present context it  
has also produced the remarkable prediction that 
there will be a component of the EMC effect 
which is isovector in nature if one 
has a target with $N \neq Z$~\cite{Cloet:2009qs}. 
Most importantly, because the EMC effect involves a 
change in the structure of the bound nucleon, 
that isovector EMC correction will 
persist even if one derives data for an effectively 
isoscalar nucleus by subtracting 
the contribution of the excess neutrons. 

In terms of the Paschos-Wolfenstein relation and the NuTeV determination 
of $\sin^2 \theta_W$ this leads to a shift of momentum from all the $u$ 
quarks in the nucleus to all the $d$ quarks. This is the same sign as the 
effect of CSV and also reduces the NuTeV anomaly by around one standard 
deviation. Figure~\ref{fig:NuTeV} shows the result of a recent reanalysis of 
the NuTeV anomaly~\cite{Bentz:2009yy} in which this correction 
was applied to the data along with a correction 
for genuine charge symmetry violation associated 
with the mass difference between 
$u$ and $d$ quarks, as well as the effect of photon radiation.
Clearly there remains no significant discrepany 
between the corrected measurement 
and the Standard Model.

\section{Conclusion}
We have briefly reviewed the latest state of play in the analysis of what used 
to be known as the NuTeV anomaly. 
It should be clear that well understood corrections 
associated with charge symmetry violation and the iso-vector EMC effect 
unambiguously remove any significant 
deviation from the Standard Model. The one major 
remaining uncertainty concerns the possible asymmetry between the strange and 
anti-strange PDFs~\cite{Signal:1987gz}. 
Even though the model independent behaviour of these 
distributions under chiral symmetry suggests that there must be some 
asymmetry~\cite{Thomas:2000ny}, it is experimentally very 
poorly determined. Working toward a precise 
measurement of $s(x) - \bar{s}(x)$ should be a very high experimental 
priority for a number of reasons, including its relevance to the NuTeV 
experiment. However,  on grounds that are admittedly somewhat model 
dependent, we find it extremely unlikely that this asymmetry 
could be large enough to make a large contribution. In particular, there is no 
known perturbative QCD mechanism to produce a large asymmetry. In terms 
of non-perturbative mechanisms, 
chiral effects are controlled by factors like 
$m_K/m_N$ and $m_\Lambda/m_N$ and hence any 
change in sign in $s(x) - \bar{s}(x)$ 
is unlikely to occur at values of Bjorken $x$ much below 0.1. 
With such a constraint 
the NuTeV analysis of their own di-muon data yields a 
tiny value for the $s-\bar{s}$ 
asymmetry~\cite{Londergan:2009kj}.

Of course, while the theoretical results presented 
here are compelling, as always 
it will be important to carry out careful experimental investigations 
of CSV of the $u$ and $d$ PDFs as well as the 
iso-vector EMC effect. Further lattice simulations 
of the moments of $s-\bar{S}$, 
complemented by dedicated experimental studies are also vital.
   
%%%%%%%%%%%%%%%%%%%%%%%%%%%%%%%%%%%%%%%%%%%%%%%%
%% BACKMATTER
%%%%%%%%%%%%%%%%%%%%%%%%%%%%%%%%%%%%%%%%%%%%%%%%

\begin{theacknowledgments}
This work was supported by the University of Adelaide and 
by the Australian Research Council through the award 
of an Australian Laureate Fellowship.
\end{theacknowledgments}


%\bibliography{csymrefR}

\end{document}